\documentclass[aps,preprintnumbers,prd,twocolumn,superscriptaddress]{revtex4}

\usepackage{epsfig,latexsym,cancel,amssymb,amsmath}
\usepackage{graphicx}
\usepackage{epstopdf}
\usepackage{hyperref, graphicx, slashed, xcolor, amsmath}

\allowdisplaybreaks

\definecolor{red}{rgb}{1,0,0}

\newcommand{\beq}{\begin{equation}}
\newcommand{\eeq}{\end{equation}}
\newcommand{\bea}{\begin{eqnarray}}
\newcommand{\eea}{\end{eqnarray}}
\newcommand{\nn}{\nonumber\\}

\begin{document}

\title{Direct detection of freeze-in inelastic dark matter}

\author{Haipeng An}
\affiliation{Department of Physics, Tsinghua University, Beijing 100084, China}
\affiliation{Center for High Energy Physics, Tsinghua University, Beijing 100084, China}
\author{Daneng Yang}
\affiliation{Department of Physics, Tsinghua University, Beijing 100084, China}

\begin{abstract}
We show that the current sensitivities of direct detection experiments have already reached the interesting parameter space of freeze-in dark matter models if the dark sector is in the inelastic dark matter framework and the excited dark matter state is cosmologically stable. Using results recently presented by the XENON1T experiment, we present constraints on these models. We also show that these models can explain the reported excess in the electron recoil signals if the mass gap between the ground state and the excited state is at keV scale.
\end{abstract}

\maketitle

\section{Introduction}

The particle nature of dark matter (DM) is one of the most prominent mysteries. Till now, all the evidence of the existence of DM is from gravitational effects. The relic energy density of DM in today's universe is measured to be about one-quarter of the total energy density. A successful DM model must be able to provide a mechanism to understand this number. The freeze-in scenario of DM production provides such a mechanism~\cite{Hall:2009bx}. In this scenario, the DM particles live in the dark sector, very weakly connecting to the standard model (SM) sector through a portal. 
It is assumed that after inflation, only the SM sector is reheated, and through the portal, the energy in the SM sector leaked into the dark sector. In this scenario the observed relic density of DM can be nicely produced. However, the direct detection channel in freeze-in models is also proportional to the portal and, therefore, strongly suppressed. Inelastic DM models were first introduced to explain the excess observed in the DAMA/LIBRA experiment~\cite{TuckerSmith:2001hy,Chang:2008gd}, since an enhanced annual modulation can be generated due to the extra cost of the kinetic energy in the up-scattering process. The down-scattering process in inelastic DM models is usually ignored since the population of the excited state is usually exponentially suppressed. In this work, we consider freeze-in inelastic DM models. We show that due to the possibility of the large down scattering rate, the sensitivity of the XENON1T experiment~\cite{Aprile:2020tmw} has already achieved the interesting parameter space of inelastic freeze-in models. In Ref.~\cite{Aprile:2020tmw}, an excess of around 1$\sim$5 keV in electron recoil events is also reported, which cannot be accounted for by known backgrounds. Since the report of this excess, there have been active investigations trying to under it with new  physics models~\cite{Bloch:2020uzh,Chala:2020pbn,Lindner:2020kko,Budnik:2020nwz,Gao:2020wer,Zu:2020idx,An:2020bxd,Baryakhtar:2020rwy,Bramante:2020zos,Jho:2020sku,Gelmini:2020xir,Nakayama:2020ikz,Primulando:2020rdk,Khan:2020vaf,Cao:2020bwd,Robinson:2020gfu,Lee:2020wmh,Paz:2020pbc,Choi:2020udy,AristizabalSierra:2020edu,Buch:2020mrg,Bell:2020bes,Dey:2020sai,Chen:2020gcl,DiLuzio:2020jjp,Du:2020ybt,Su:2020zny,Harigaya:2020ckz,Bally:2020yid,Boehm:2020ltd,Fornal:2020npv,Amaral:2020tga,Alonso-Alvarez:2020cdv,Kannike:2020agf,OHare:2020wum,Takahashi:2020bpq}. In this work we show that this excess can be explained in the framework of inelastic freeze-in models. 

\bigskip

\section{Vector portal freeze-in models}

In this paper, we consider two typical models: one with a complex scalar DM and the other with a Dirac spinor DM. In both models, the dark portal is assumed to be a vector field $V$, which we call a dark photon in the following discussion. In both cases, we assume the U(1) symmetry is broken softly by an explicit mass splitting. In the scalar case, the masses of the real and imaginary parts are split, and in the Dirac spinor case, the spinor is split into two Majorana spinors. The DM part of Lagrangian can be written as
\bea\label{eq:La}
{\cal L}^{sc} &=& \frac{1}{2} \partial_\mu \chi_1 \partial^\mu\chi_1 + \frac{1}{2} \partial_\mu \chi_2 \partial^\mu\chi_2 - \frac{1}{2} m_1^2 \chi_1^2  - \frac{1}{2} m_2^2 \chi_2^2 \nn
&& - e_D V^\mu (\chi_1  \partial_\mu \chi_2 - \chi_2 \partial_\mu \chi_1 ) + \frac{1}{2} e_D^2 V_\mu V^\mu (\chi_1^2 + \chi_2^2), \nn
{\cal L}^{sp} &=& \chi_1 ^\dagger i\sigma^\mu \partial_\mu \chi_1 + \chi_2 ^\dagger i\sigma^\mu \partial_\mu \chi_2 - \frac{1}{2} (m_1 \chi_1 \chi_1 + m_2 \chi_2 \chi_2\nn &&+ {h.c.})
 + i e_D V^\mu (\chi_1^\dagger \sigma_\mu \chi_2 - \chi_2 ^\dagger \sigma_\mu \chi_1), \nn
\eea
where $\chi_1$, $\chi_2$ are real components of a complex scalar $\chi$ in ${\cal L}^{sc}$, 
and two-component Weyl spinors for a pseudo Dirac fermion $\chi$ in ${\cal L}^{sp}$. 


The Lagrangian for the dark photon part can be written as 
\bea
{\cal L}^V = -\frac{1}{4} V_{\mu\nu} V^{\mu\nu} + \frac{1}{2}{m_V^2}{V_\mu V^\mu} - \frac{\kappa}{2} V_{\mu\nu} F^{\mu\nu} \ .
\eea

To avoid having too much down-scattering through dark photon exchange processes, one can introduce extra interactions to moderate the rate.
In the scalar case, it is natural to consider a term ${\cal L}_\lambda = -\dfrac{\lambda}{8} \left( \chi_1^2 + \chi_2^2 \right)^2$. For the sake of vacuum stability, without inducing other interactions in the potential, $\lambda$ must be positive. Then the combined square of the absolute value of the matrix element for the down scattering in the NR limit becomes
\bea
|{\cal M}|^2_{\chi_2\chi_2  \rightarrow \chi_1 \chi_1} = \left| \frac{8 e_D^2 m_D^2}{m_V^2} - \lambda\right|^2 \, 
\eea
resulting in a suppression in the down-scattering rate. 

In the spinor case, one can introduce a scalar-carried force in the dark sector with a term ${\cal L}_h = -\dfrac{y}{2} \left( h \chi_1\chi_1 + h \chi_2\chi_2 + {h.c.}\right)$~\footnote{Here, we only introduce the diagonal interaction of $h$ to $\chi_1$ and $\chi_2$ because the off-diagonal coupling is prevented by a $Z_2$ symmetry, defined as $Z_2(V) = Z_2(\chi_1) = -1$, and $Z_2(\chi_2) = Z_2(h)=1$. This symmetry is indeed inherited from the $C$ parity of Dirac spinors. }. 
The down-scattering amplitude through this interaction is $s$-channel, whose sign depends on the mass of $h$. 
By choosing sign different from the $V$ exchange amplitude, one is arrived at a similar cancellation as in the scalar case. 

The excited states $\chi_2$ may also decay into the ground state $\chi_1$. 
As long as $\Delta \equiv m_2 - m_1$ is smaller than $m_V$ and $2 m_e$, the dominant channels for this decay are the three-photon channel and the neutrino channel. The former is suppressed by a factor of $\kappa^2 (\Delta/m_e)^8$, and the latter is suppressed by $\kappa^2 (m_V/m_Z)^4$. Therefore, the lifetime of $\chi_2$ can be much longer than the age of the universe. For the purpose of direct detection we assume $\Delta \ll m_1 \approx m_2$.

\bigskip

\section{Freezing-in the DM}

\begin{figure*}[t!]
        \includegraphics[scale=0.33]{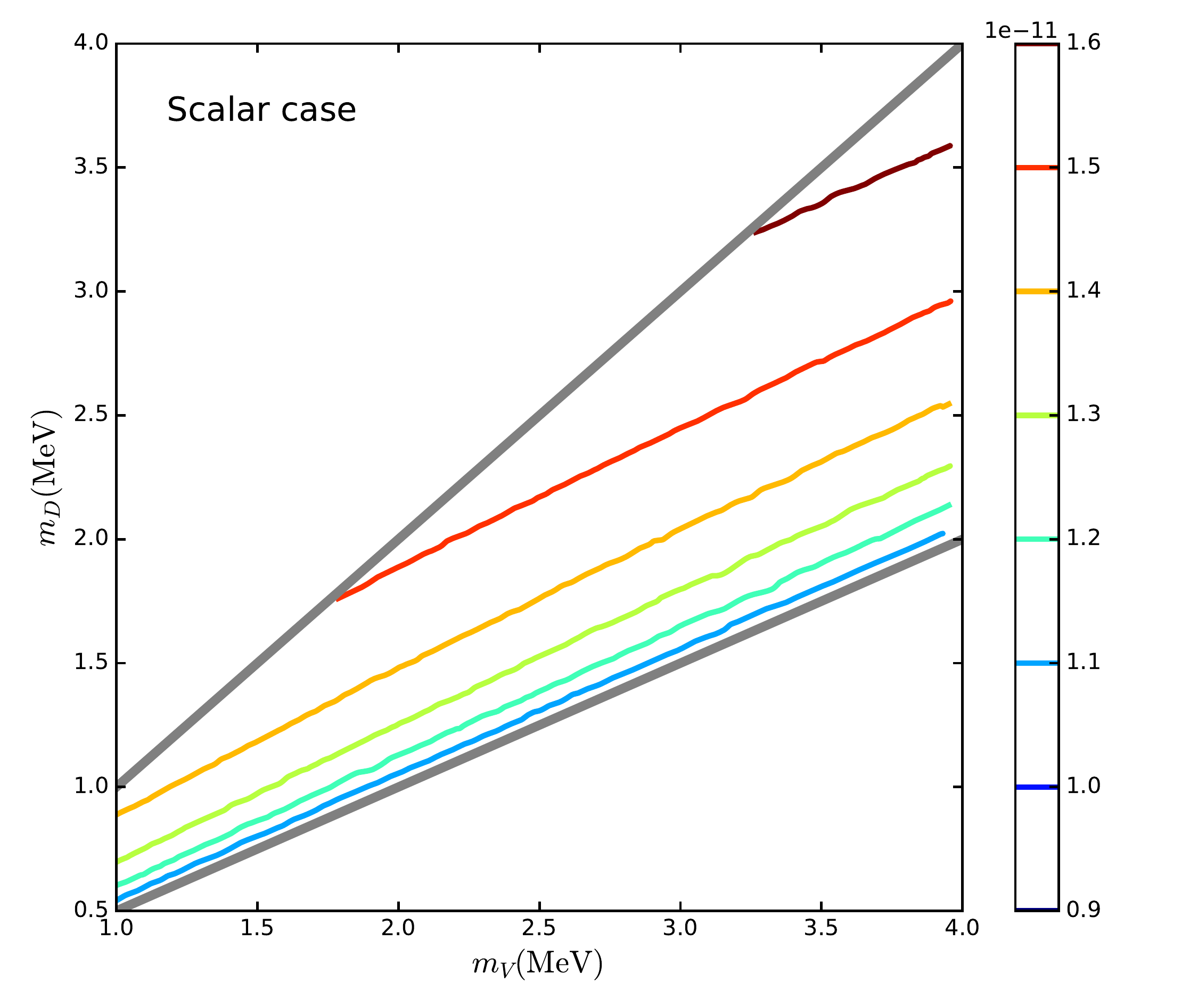}
        \includegraphics[scale=0.33]{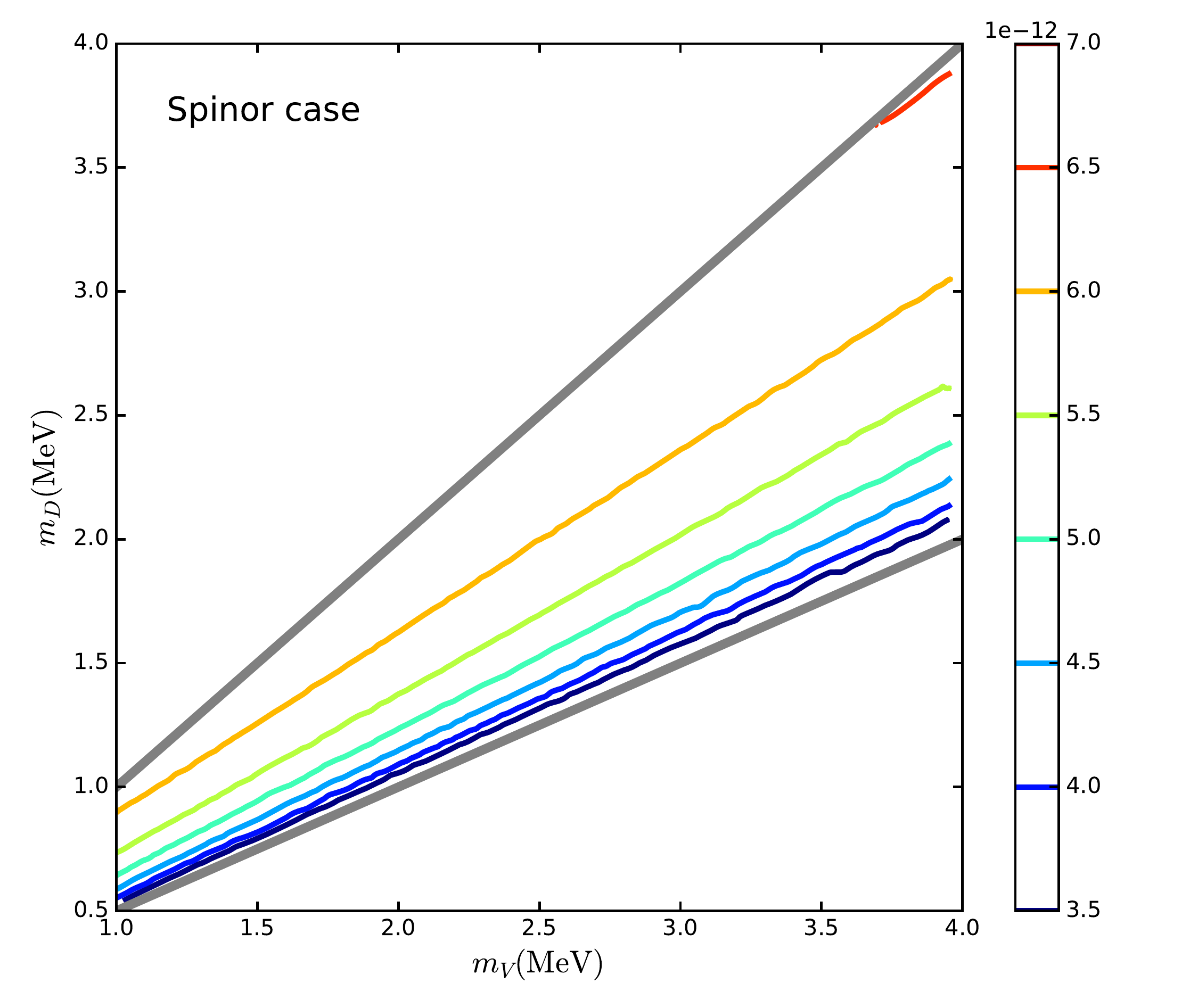}
\caption{Size of $\kappa e_D$ in scalar (left) and spinor (right) DM models required to produce the relic DM abundance.  }
\label{fig:freezein}
\end{figure*}

To calculate the relic density of DM we need to solve the Boltzmann equation
\bea\label{eq:boltzmann1}
\frac{d n_D}{d t } + 3 H n_D = \Gamma_{fi} \ ,
\eea
where $n_D$ is the DM number density, $H$ is the Hubble parameter, and $\Gamma_{fi}$ is the production rate of DM per volume. When the 	temperature $T_{\rm SM} > $ 1 MeV, the universe is filled with relativistic plasma, therefore a $\Gamma_{fi}$
can be estimated as $\kappa^2 e_D^2 \alpha_{\rm em} T_{\rm SM}^4$. The time interval at certain $T_{\rm SM}$ can be estimated as $H^{-1}\sim {m_{\rm pl}/T_{\rm SM}^2}$, where $m_{\rm pl} \approx 1.22\times10^{19}$ GeV, is the Planck mass. Then the produced number density per entropy can be written as
$y_D \equiv n_D/s \sim \kappa^2 e_D^2 \alpha_{\rm em} \times (m_{\rm pl}/T_{\rm SM})$. Therefore in this scenario, DM is mainly produced at low temperate. The relation between $y_D$ and $T_{\rm SM}$ stops when $T_{\rm SM}$ hits either $m_V$, $m_D$ or $m_e$. 
For $m_e < m_V <  2 m_D$, which will be motivated later, the 
freeze-in process stops at $T_{\rm SM}\sim m_D$. Consequently, the dependence of today's relic density $\Omega_D\propto y_D m_D$ on $m_D$ is canceled. As an order of magnitude estimation, we can roughly get 
\bea
\Omega_D \sim \kappa^2 e_D^2 \alpha_{\rm em} \times \frac{m_{\rm pl}}{m_{p}  \eta_\gamma} \ ,
\eea 
where $m_p$ is the proton mass. As a result, to get the observed relic abundance, the product of the dark coupling $e_D$ and kinetic mixing $\kappa$ is fixed to $\kappa e_D \sim 10^{-13}\sim 10^{-12}$. The numerical results of $\kappa e_D$ required to produce the observed relic abundance for different choices of $m_V$ and $m_D$ are shown in Fig.~\ref{fig:freezein}. The direct detection rate is also proportional to the factor $\kappa^2 e_D^2 \alpha_{\rm em}$, making it difficult to search for the freeze-in model. It is also proportional to the number density of the DM, and therefore in favor of low mass DM as long as the energy deposit can surpass the thresholds of the experiments. Therefore in this work, we focus on the region where both $m_D$ and $m_V$ are around MeV scale. 

In this regime, there are two main contributions to $\Gamma_{fi}$: one is through $e^+ e^-$ annihilation, the other is through plasmon decay. The details of these two processes can be found in Ref.~\cite{An:2018nvz} (see also \cite{Dvorkin:2019zdi}). In our case, since we require that $m_V < 2 m_D$ the freeze-in cannot go through on-shell $V$, and the contribution from plasmon decay is always subdominant (The similar phenomenon is also found in the freeze-in process of on-shell dark photon~\cite{Fradette:2014sza}). 

For the e+ e- annihilation contribution, we have
\bea
\Gamma_{fi}^{sc} &\approx& \frac{\kappa^2 e_D^2 \alpha_{\rm em} }{24\pi^3} \int dq \int dq^0 f\left( \frac{q^0}{T},\frac{q}{T}, s \right)  \nn
&&\times \frac{q^2 (s+2m_e^2)s(1-4m_D^2/s)^{3/2}}{(s-m_V^2)^2} \ ,\nn
\Gamma_{fi}^{sp} &\approx& \frac{\kappa^2 e_D^2 \alpha_{\rm em} }{6\pi^3} \int dq \int dq^0 f\left( \frac{q^0}{T},\frac{q}{T}, s \right)  \nn
&&\times \frac{q^2 (s+2m_e^2)(s+2m_D^2)(1-4m_D^2/s)^{1/2}}{(s-m_V^2)^2}\ ,
\eea
where the superscripts $sc$, $sp$ denote the scalar and spinor cases, respectively. 
The function $f(x,y,s)$ is defined as: 
\bea
f(x,y,s) = \frac{1}{2\pi y} \frac{4~ {\tanh}^{-1}\left[ \left(\frac{a-1}{a+1}\right)\tanh\left(\frac{b}{2}\right) \right]}{(a-1)(a+1)} \ ,
\eea
with
$a = e^{x/2}$ and $b = \frac{y}{2} \left( 1 -  \frac{4m_e^2}{s}\right)^{1/2}$. The difference between between $\Gamma^{sc}_{fi}$ and $\Gamma^{sp}_{fi}$ is due to the difference of the spin structure and the factor of $(1- 4m_D^2)^{3/2}$ is due to the p-wave nature of the decay of virtual $V$ into scalars. Here and the following we use $m_D \approx m_1 \approx m_2$ in the calculation when $\Delta$ can be neglected. 

\bigskip

\section{Direct detection rates and recoil spectrum}

When $\chi_2$ particles fly into the XENON detector, the ionization process through the down-scattering of $\chi_2$ can happen. 
To calculate the ionization rate, we assume the scattering electron is approximated by
a plane wave; the xenon atom is isolated and described by the Roothaan-Hartree-Fock ground state wave functions~\cite{Bunge:1993jsz}.
It follows that the velocity averaged differential ionization cross section times velocity for electrons in the $(n.l)$ shell can be written as
\bea\label{eq:11}
\frac{d \langle \sigma^{n l}_{\rm ion} v\rangle }{d \ln E_r} &=& \frac{\kappa^2 e_D^2 \alpha_{\rm em}}{2 m_V^4} F^{nl}(E_r) \int q dq \left\langle \frac{1}{v}\theta(v - v_{\rm min})\right\rangle \nn
\!\!\!\!\!\!&&\!\!\!\!\!\!\times \frac{k^3}{(2\pi)^3} \int d\Omega_k \left| \int d^3x e ^{- i \vec k \cdot \vec x- i \vec q \cdot \vec x} \psi_{nlm} \right| \ ,
\eea
where $E_r$ and $k$ are the kinetic energy and momentum of the outgoing electron, $q$ is the momentum-transfer, and $v$ is the velocity of the incoming $\chi_2$. 
The function $F^{nl}(E_r)$ is an enhancement factor induced by the attractive potential around the nucleus. We follow Ref.~\cite{Bloch:2020uzh} and consider 
\begin{eqnarray}
F^{nl}(E_r) &=& \dfrac{2\pi \zeta^{nl}}{1-e^{-2\pi \zeta^{nl}}}, 
\end{eqnarray}
where $\zeta^{nl} = Z_{eff}^{nl} \alpha_{\rm em} \sqrt{m_e/(2 E_r)}$ and $Z_{eff}^{nl} = \{12.4, 14.2, 21.9, 25.0, 26.2, 39.9, 35.7, 35.6, 49.8, 39.8, 52.9\}$ for the $nl$ shells
$\{5p, 5s, 4d, 4p, 4s, 3d, 3p, 3s, 2p, 2s, 1s\}$~\cite{Bloch:2020uzh,doi:10.1063/1.1733573,doi:10.1063/1.1712084}.
According to the principles of quantum mechanics, scattering states and bound states from the same Hamiltonian must be orthogonal. Therefore, in the domain that $|\vec q \cdot \vec x| \lesssim 1$, the plane wave approximation overestimates the cross section. 
To avoid this spurious contribution, we subtract the bound state component from the outgoing wave function:
\bea
e^{i \vec k \cdot \vec x} \rightarrow e^{i \vec k \cdot \vec x} - \int d^3y e^{i \vec k \cdot \vec y} \psi_{nlm}^*(y) \psi_{nlm}(x) \ .
\eea
The form factors calculated in this way agree reasonably well with the ones used in~\cite{Essig:2017kqs}. In the case of down-scattering, 
\bea
v_{\rm min} = \left| \frac{E_r}{q} + \frac{q}{2 m_D} + \frac{E_B - \Delta}{q} \right| \ ,
\eea
where $E_B$ is the absolute value of the binding energy. If $E_B \ll \Delta$, the electrons indeed can be treated as free particles, and one can estimate $ \sigma v $ as follows without going through the complicated form factor evaluations:  
\bea
\left(\sigma v\right)_{\rm ion}^{tot} \approx \sum_{|E_B| < \Delta m} 4\sqrt{2} \frac{\kappa^2 e_D^2 \alpha_{\rm em} \mu^2}{m_V^4} \left(\frac{\Delta}{\mu}\right)^{1/2} \ ,
\eea
where $\mu = m_e m_D /(m_e + m_D)$, and the summation is over all the orbits with binding energy smaller than $\Delta$.  

\begin{figure}
\centering
\includegraphics[height=2.0in]{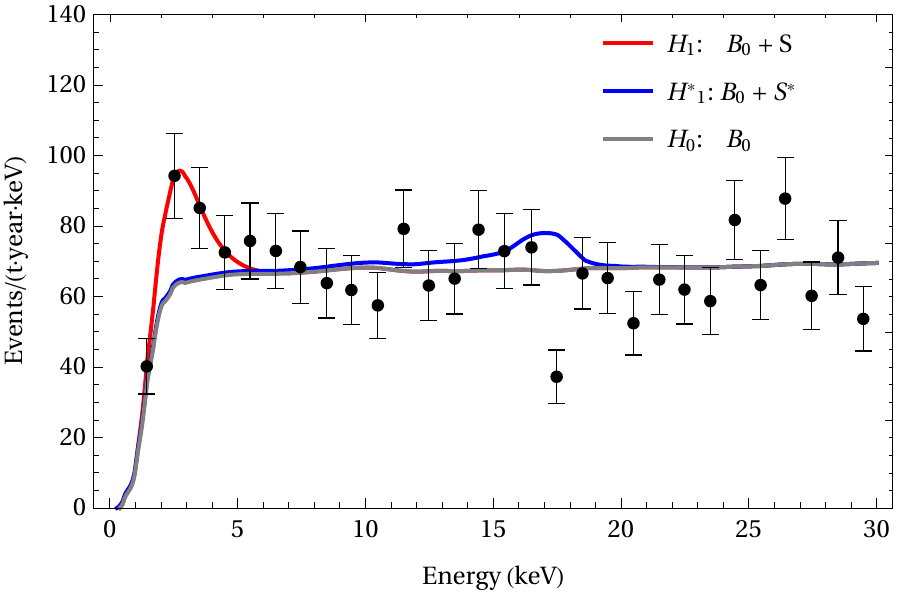}
\caption{Down-scattering detection signals from freeze-in inelastic DM model.
The red curve is obtained from a fit to data using the background template in Ref.[\cite{Aprile:2020tmw}] and a signal spectrum with $m_D=0.8~$MeV, $\Delta=5~$keV and a floating $f_2$.
The blue curve represents a signal with $m_D=1.3~$MeV, $\Delta=24~$keV that corresponds to a 95\% CL limit shown in Fig.(\ref{fig:contCL}).
}\label{fig:spectrum}. 
\end{figure}

\begin{figure}
\centering
\includegraphics[height=2.0in]{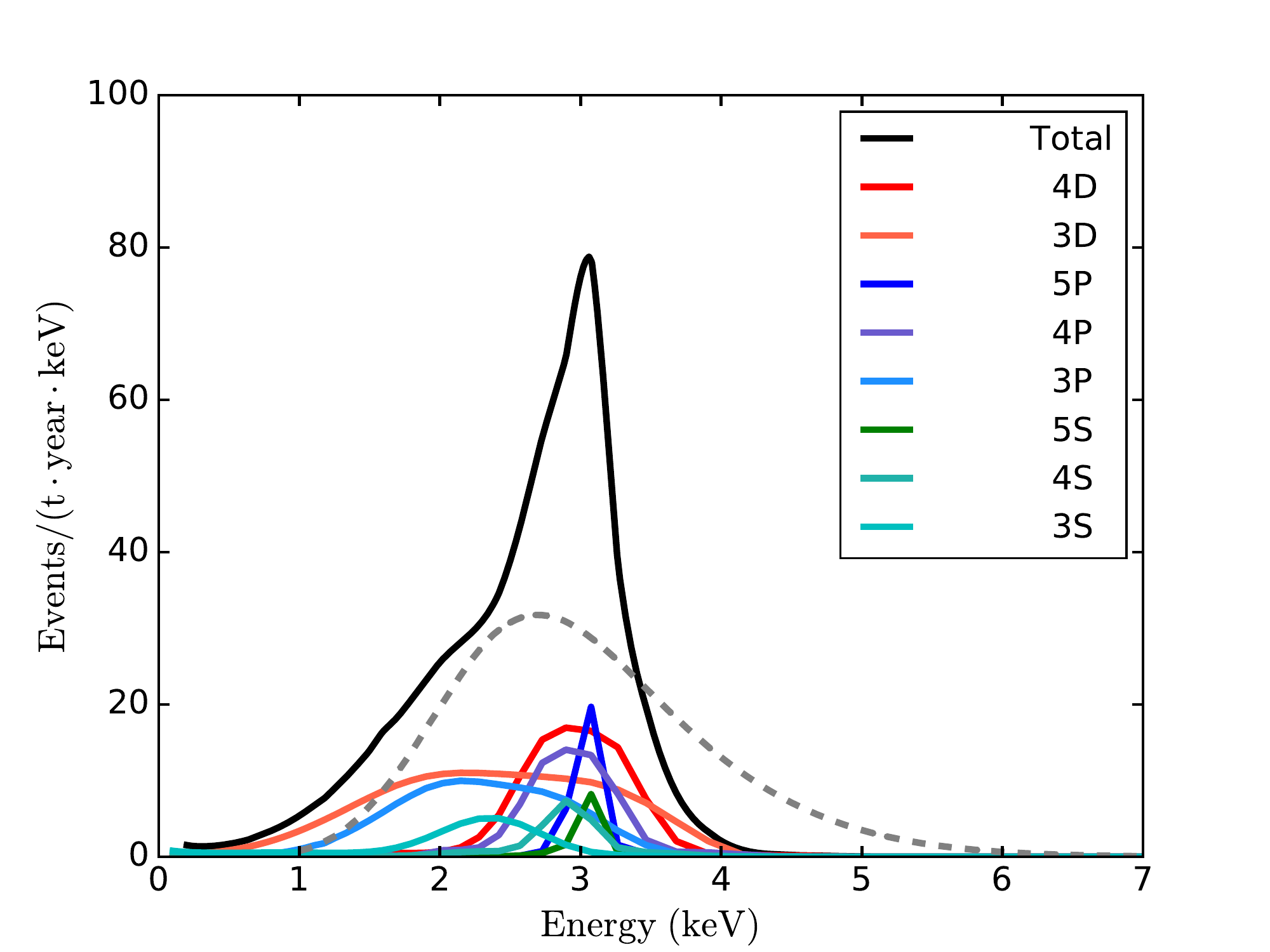}
\caption{Total recoil energy spectrum (Black) and its composition for $m_D = 0.8$ MeV, $m_V = 1.2$ MeV, $\kappa e_D = 1.29\times 10^{-11}$, and $\Delta = 5$ keV.
Each colored curve corresponds to a contribution from an energy level. The total expected observation, taking into account effects of detector resolution and efficiency, is shown as a dashed gray curve. }\label{fig:partial}.
\end{figure}


The differential ionization rate in the detector can be written as
\bea
\frac{d R_{\rm ion}}{d \ln E_r} = f_2 N_T  \frac{\rho_D}{m_D} \frac{d \langle \sigma_{\rm ion}^{\rm tot} v\rangle}{d \ln E_r } \ ,
\eea
where $N_T$ is the total number of the target atoms, $f_2 \equiv n_2/n_D$ is the fraction of the excited state. 
The value of $f_2$ strongly depends on the detailed model of the dark sector
and we decide to discuss the physics in a latter section. 
Here we choose different values of $f_2$ and calculate the constraint on $\Delta$, $m_V$ and $m_D$. The electron recoil spectra on top of the known background of the XENON1T experiment are shown in Fig.~\ref{fig:spectrum}. For the red curve we choose $m_D = 0.8$ MeV, $m_V = 1.2$ MeV, $\kappa e_D = 1.29\times 10^{-11}$, $\Delta = 5$ keV and extract $f_2 = 5.54\times 10^{-3}$ from a fit. For the blue curve, we choose $m_D = 1.3$ MeV, $m_V = 2.0$ MeV, $\kappa e_D = 1.35\times 10^{-11}$, $\Delta = 24$ keV and $f_2 = 0.02$. The red curve is the best fit signal for the excess and the blue one corresponds to a 95\% CL exclusion. The parameters for both the two curves can produce the observed relic abundance. We can see that with the XENON1T data we can already put constraints on the parameter space of the model. 

We show in Fig.~\ref{fig:partial} the contributions from each energy level to the recoil spectrum. 
In this study, $\Delta \gg E_B$ and the DM particles have negligible kinetic energies. 
Based on kinematical considerations, one expect the spectrum to peak around $m_D \Delta/(m_e + m_D)$ with $|\vec q + \vec k|\sim 0$. 
From Eq.~(\ref{eq:11}), we found the form factor is peaked when $|\vec q + \vec k| \lesssim 1/r_n$, where $r_n$ is the size of the bound state wave function on the $n^{th}$ shell. 
Therefore, the size of the bound state wave function, which may be estimated using the Bohr model, $r_n\approx n^2/(Z_{\rm XENON} \alpha_{em} m_e)$, leads to a width in the theoretical recoil spectrum. 
Let $|\vec{k} + \vec{q}| = |\delta\vec{k}| \approx 1/r_n$, one obtains $\delta E_r\approx |\vec{k}\delta\vec{k}|/(m_e) \approx \alpha_{em} Z_{\text{XENON}} \sqrt{2\mu\Delta}/n^2$. 
The $n$ dependence of the width is clear in Fig.~\ref{fig:partial}. 
For electrons ionized from the layers 4S, 3D, 3P, and 3S, the peaks of the $E_r$ distributions are shifted because of the non-negligible binding energies.
In practice, as shown by the dashed gray curve, the width of the measured recoil spectrum is mainly determined by the detector energy resolution.

\begin{figure}
\centering
\includegraphics[height=2.8in]{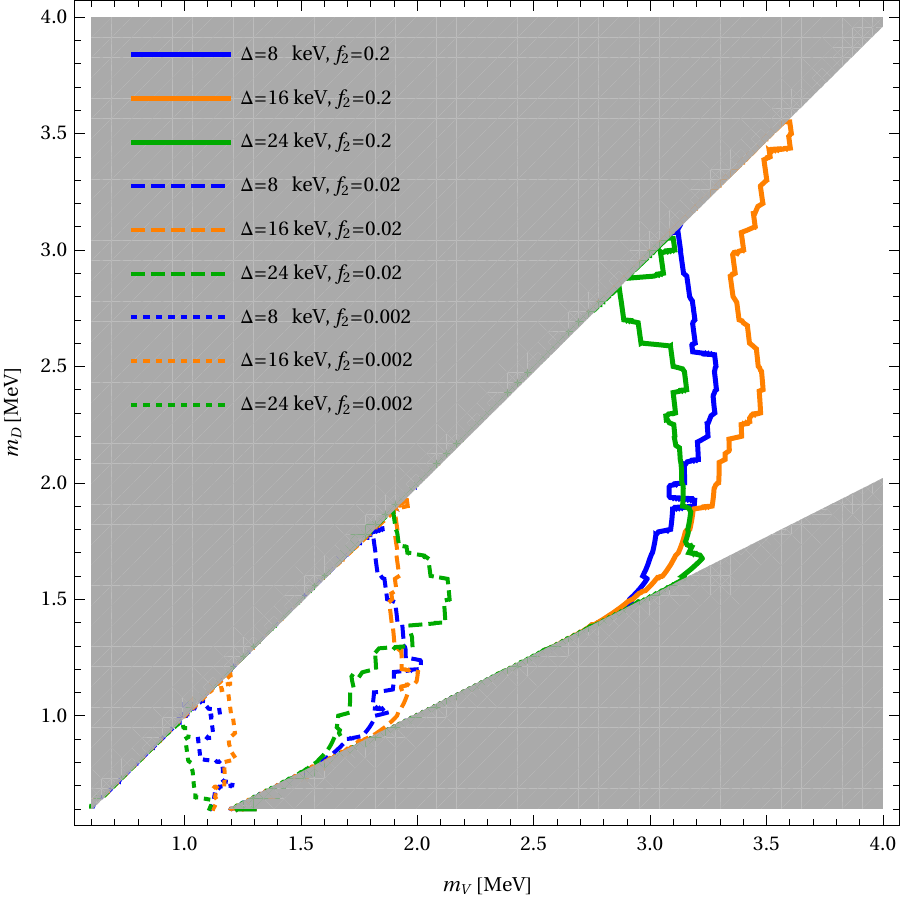}
\caption{Observed 95\% CL limits in the parameter space of the inelastic freeze-in model with scalar DM and different $\Delta$, $f_2$. The excluded regions lie on the lower-left. }\label{fig:contCL}.
\end{figure}

We perform statistical analysis assuming the same test statistic as in Ref.\cite{Aprile:2020tmw} but without systematic uncertainties, which are neglectable comparing to the statistical ones.
In the high energy sideband of the spectrum where there is no observable excess, we use the XENON1T data to constrain the model parameter space.
Using the asymptotic distribution of the test statistic in the large sample limit, we use the experiment spectrum in a range $E_R=\frac{m_D \Delta}{m_D+me}\pm 2\Delta E_R$ to set observed 95\% CL limit on the signal strength, where the $\Delta E_R$ is the detector resolution extracted from Ref.\cite{XENON:2019dti}.
We fix $\kappa e_D$ to reproduce the relic DM abundance in the scalar case, whose value has been shown in Fig.(\ref{fig:freezein}).
The exclusion contours using the XENON1T spectrum are shown in Fig.(\ref{fig:contCL}) at fixed $\Delta$ and $f_2$.
Regions on the lower-left are excluded, and the constraints are stronger in regions of larger $\Delta$ and $f_2$.

\begin{figure}
\centering
\includegraphics[height=2.6in]{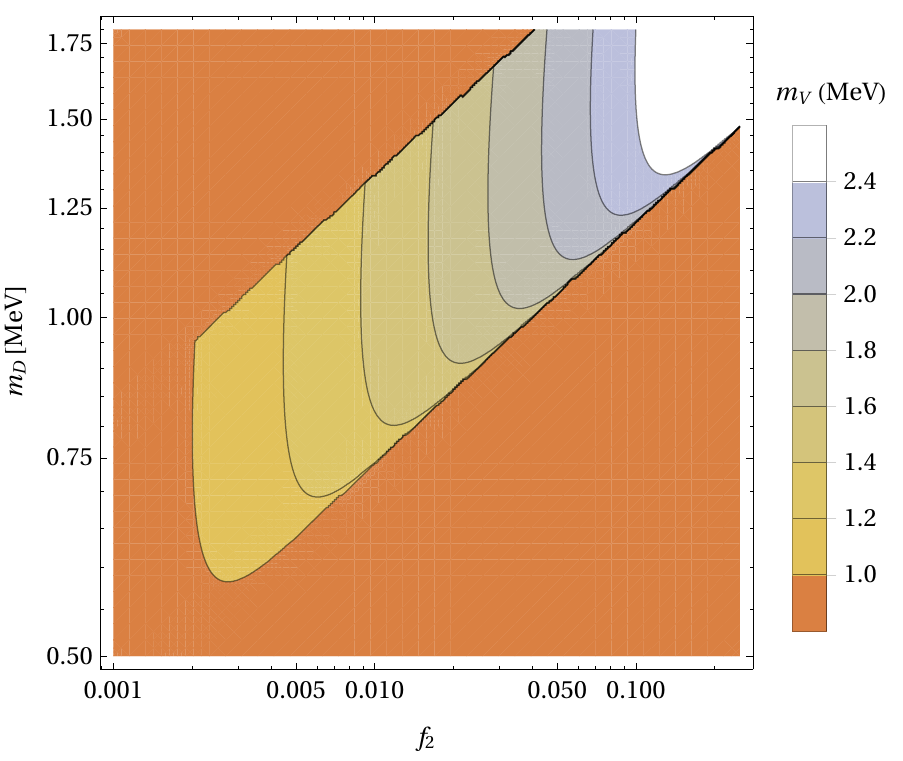}
\caption{Parameter space of the inelastic freeze-in model with scalar DM that is consistent with the best fit XENON1T excess, assuming no tritium contribution in the background template. The orange region is inconsistent with the requirement $m_D<m_V<2 m_D$. }\label{fig:excess}.
\end{figure}

In the low energy region with the observable excess, we fit the XENON1T data using our simulated signal spectra. For the benchmark point in Fig.(\ref{fig:spectrum}), the best fit signal with $f_2=5.54\times 10^{-3}$ is found to be $3.6\sigma$ favored over the background-only hypothesis, which decreases to $2.6\sigma$ if the shape of tritium contribution is included as an unconstraint component.
To find the parameter region consistent with this excess, we consider spectra whose $E_R \approx 2.7~$keV, fix $\kappa e_D$ as in the scalar case of Fig.(\ref{fig:freezein}), and adjust $f_2$ such that the yield equals to that of the best fit benchmark. Results of the parameter scan are shown in Fig.(\ref{fig:excess}) as contours of $m_V$. We found that the model in this work can provide viable solutions to the XENON1T excess for dark photon masses between one and two MeV.
The contour plot with different choices of $m_V$ from $2 m_e$ to about $2.5$ MeV to fit the excess is also shown in Fig.~\ref{fig:excess}, one can see that to get enough number events $f_2$ must be larger than a few $10^{-3}$.

\bigskip

\section{De-excitation in the early universe}

\begin{figure}
\centering
\includegraphics[height=1.5in]{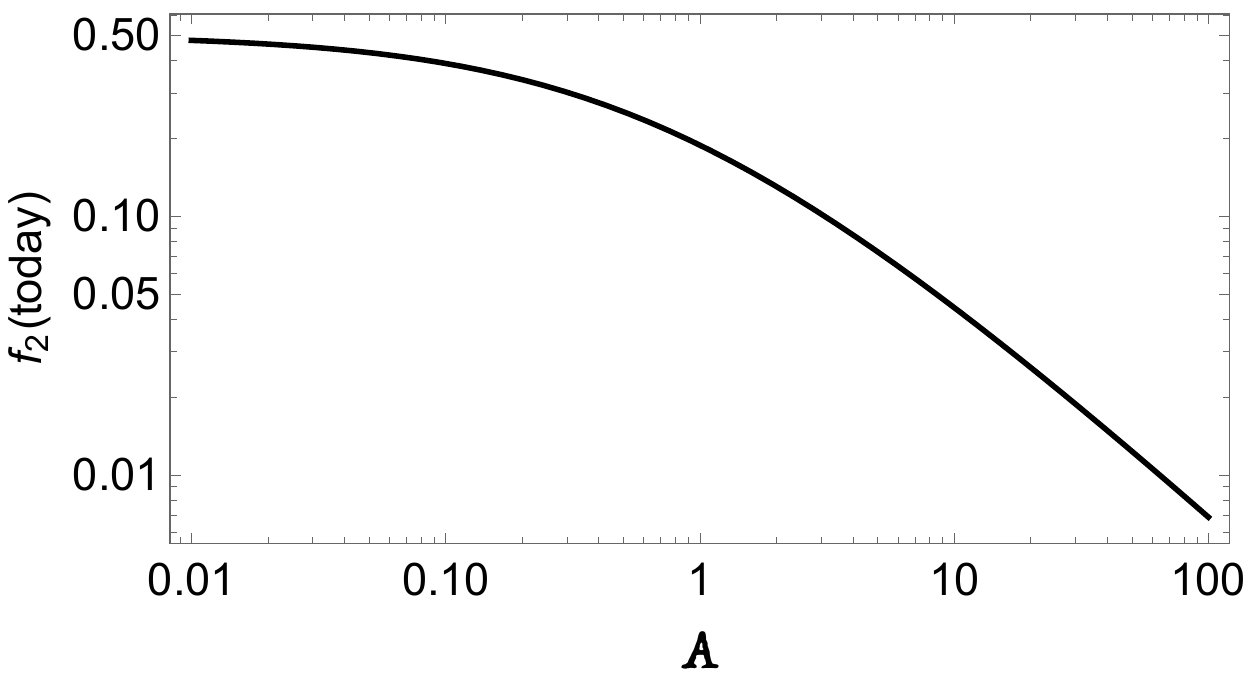}
\caption{Today's fraction of $\chi_2$ as a function of ${\cal A}$ defined in Eq.~(\ref{eq:A}).}\label{fig:frac1}.
\end{figure}

Although $\chi_2$ is cosmologically stable, the $\chi_2 \chi_2 \rightarrow \chi_1 \chi_1$ in the early universe can de-excite $\chi_2$ into $\chi_1$. During freeze-in the temperature or ``average kinetic energy'' $T_D$ is comparable to $T_{\rm SM}$ and is roughly equal to $m_D$.
As $\Delta \ll m_D$, the occupation of $\chi_1$ and $\chi_2$ are almost equal to each other.
During expansion of the universe, $T_D$ evolves non-relativistically. As a result we can parameterize $T_D$ as $T_D = \eta T^2 / m_D$, where $\eta \sim {\cal O}(1)$ can be determined by solving the Boltzmann equation. The Boltzmann equation for $n_1$ at $T_D \sim \Delta \ll m_D \sim m_V$ can be written as
\bea
\frac{d n_2}{d t} + 3 H n_2 = - {\cal C}(T_D) \left( n_2^2 e^{\Delta/T_D} - n_1^2 e^{-\Delta/T_D} \right) \ ,
\eea
where the collision coefficient ${\cal C}$, if ignoring the additional interaction terms associated with $\lambda$ or $y$, can be written as
\bea\label{eq:C}
{\cal C}^{(0)} = \frac{e_D^4 m_D^2 }{\pi m_V^4} \left( \frac{T_D}{\pi m_D} \right) \left(\frac{\Delta}{T_D}\right) K_1 \left(\frac{\Delta}{T_D}\right) \
\eea
for both the scalar and spinor models introduced in Eq.~(\ref{eq:La}).
Here $K_1$ is the modified Bessel function with index 1.
The decoupling of the de-excitation happens during $T_D \sim \Delta \ll T_{\rm SM}$, so the universe is still at the radiation dominated era, and therefore the Hubble expansion rate is determined by the energy density of the SM sector. Defining $x = \Delta / T_D$ we have
\bea\label{eq:f_2}
\frac{d f_2}{d x} = - {\cal A} x^{-1} K_1(x) \left[ f_2^2(x)  e^{x}- (1 - f_2)^2 e^{-x} \right] \ ,
\eea
where
\bea\label{eq:A}
{\cal A}\! &=&\! \left(\frac{9\sqrt{5}}{16\pi^4} \frac{m_{\rm pl}^3 H_0^2 \Omega_D}{g_\star^{1/2} T_{\rm CMB}^3}\right) \left( \frac{e_D^4 m_D \Delta}{ m_V^4 \eta^{1/2} } \right) \nn
\!\!&\approx&\!\frac{0.37}{\eta^{1/2}} \left(\frac{e_D}{10^{-3}}\right)^4 \left(\frac{m_V}{1.5{\rm MeV}}\right)^{-4}  \frac{m_D}{0.9{\rm MeV}}  \frac{\Delta}{ 4 {\rm keV}} \ ,
\eea
and $H_0$ is today's Hubble parameter.
The dependence of today's $f_2$ on ${\cal A}$ is shown in Fig.~\ref{fig:frac1}, where one can see that
$f_2^{\rm today} \approx 0.5$ for ${\cal A}\ll 1$, and $f_2 \approx (2{\cal A})^{-1}$ for ${\cal A}\gg 1$.

MeV scale dark photon can be copiously produced during the supernova explosion, and therefore reduce the amount of the released neutrinos, which will conflict with the observations.
For MeV scale dark photon, the constraint on the kinetic mixing can be written as~\cite{Essig:2017kqs}
\bea\label{eq:nova}
\kappa < 2.5 \times 10^{-9} ({1~\rm MeV}/m_V)^2 \ .
\label{eq:snconst}
\eea
For fixed $m_V$ and $m_D$ and by requiring all the DM particles observed today are produced from this mechanism, it can be translated to a lower bound on $e_D$. 
Take the scalar model as an example. Eq.~(\ref{eq:nova}) can be translated to $e_D > 5\times 10^{-3}$ if one takes $\kappa e_D \approx 1.25\times 10^{-11}$ according to Fig.~\ref{fig:freezein}. Then from Eqs.~(\ref{eq:f_2}) and (\ref{eq:A}) one can get that in this case
\bea
f_2 \lesssim 2\times10^{-3} \left(\frac{m_V}{1.5 {\rm MeV}}\right)^4 \left(\frac{m_D}{0.9 {\rm MeV}}\right)^{-1} \left(\frac{\Delta}{4 {\rm keV}}\right)^{-1}
\label{eqf2}
\eea
which is around the lower bound of the parameter region to explain the excess (see Fig.~\ref{fig:excess}). 
Similarly, for the spinor DM case, in the minimal model (\ref{eq:La}), $f_2$ is also marginal in fitting the excess.  
However, one can easily open up the parameter space by introducing interaction terms associated with $\lambda$ or $y$.
In the spinor case, e.g., the collision term of the Boltzmann equation of $f_2$ will be scaled as
\begin{eqnarray}
{\cal A} \rightarrow {\cal A} \left( 1 - \dfrac{\lambda m_V^2}{8 e_D^2 m_D^2} \right)^2. 
\end{eqnarray}
Moreover, in both the scalar and spinor DM cases, the newly introduced interactions contribute neither to the freeze-in process nor the direct detection,
serving only to cancel the down-scattering rate. 

\begin{figure*}[t!]
        \includegraphics[scale=0.83]{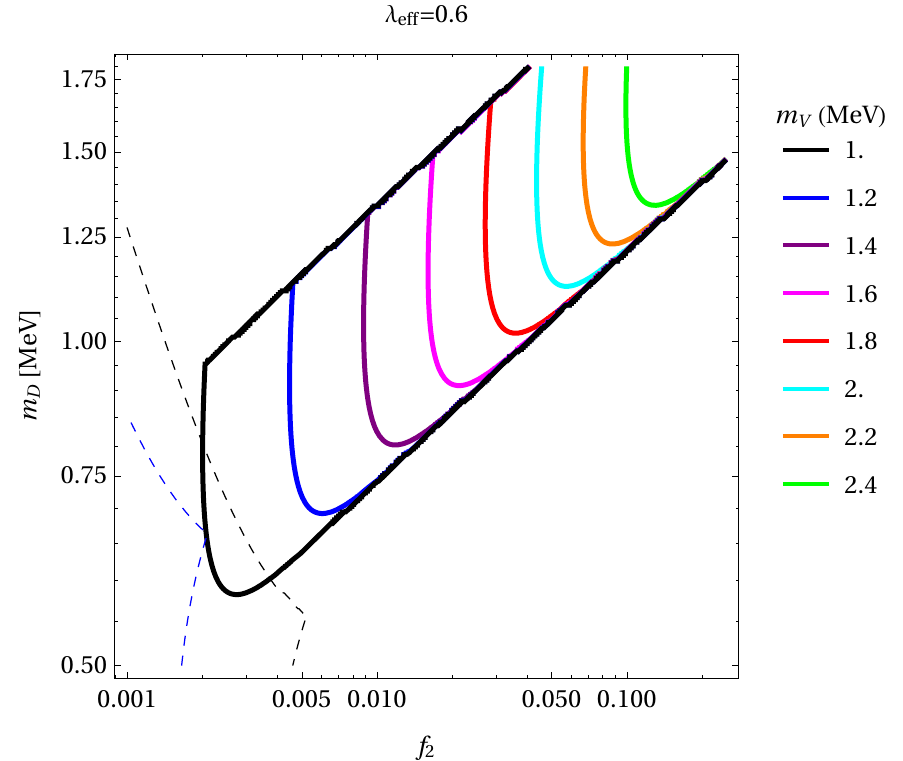}
        \includegraphics[scale=0.83]{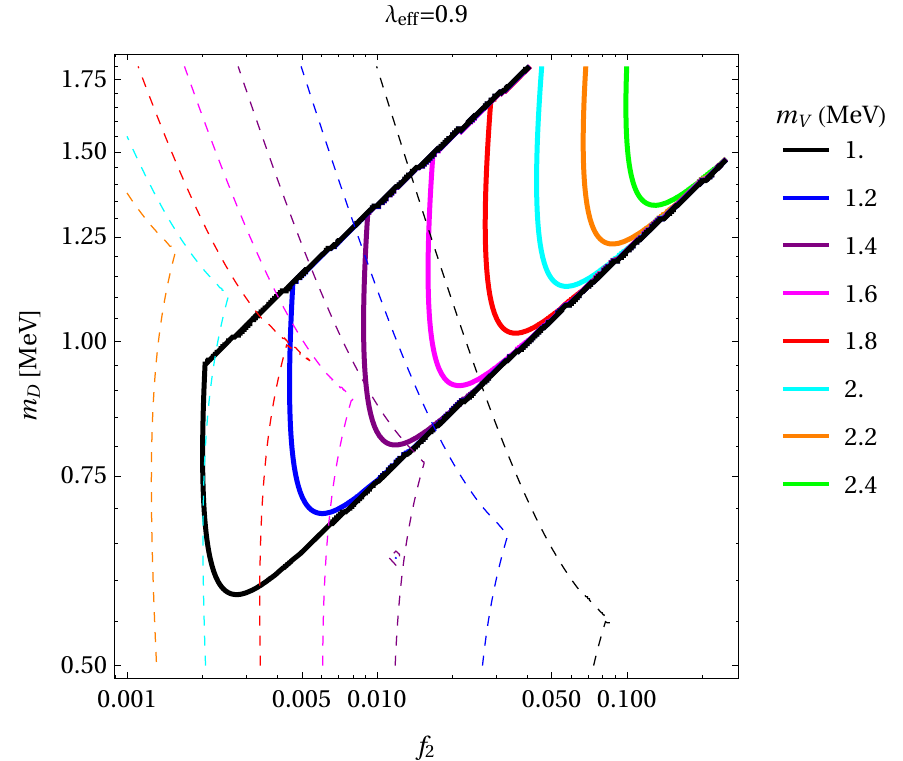}
\caption{Impact of cancellation in the down-scattering rate on the parameter space to fit the XENON1T excess, with $\lambda_{eff}=0.6$ (left) and $\lambda_{eff}=0.9$ (right).  
Colored contours in the figures are shown for the corresponding $m_V$ values that can give rise to the XENON1T excess in solid and that can saturate
the supernova constraint in Eq.(18) in dashed.
The lower-left region of the dashed contours are satisfied by this supernova constraint.}
\label{fig:cancel}
\end{figure*}

To see the impact of this cancellation on the parameter space to fit the excess, we introduce $\lambda_{eff} = \dfrac{\lambda m_V^2}{8 e_D^2 m_D^2}$ and plot in Fig.~\ref{fig:cancel}
contours of $m_V$ derived by saturating the supernova constraint Eq.~(\ref{eq:snconst}) and $f_2 = 1/(2 {\cal A})$. The new contours depicted in dashed blue lines are overlaid with the $m_V$ contours which correspond to the excess shown in solid lines. 
For the same contour level, the solid line needs to sit to the lower-left of the dashed one in order to satisfy the supernova constraint.
We found the supernova constraint can be satisfied when $\lambda_{eff} \gtrsim 0.6$.
For $\lambda_{eff} = 0.6$ shown in Fig.7 left, only $m_V$ values close to one MeV can satisfy this constraint.
For $\lambda_{eff} \gtrsim 0.9$, the range of $m_V$ is opened up to about $1.4$ MeV.

\bigskip



\section{Summary}

Freeze-in models with small couplings to the SM field is known to be challenging to search and constrain. However, in models that the DM is composed of a two-state system and with the excited state still populated in the universe, the direct detection signal can be enhanced in both the recoil energy and the rate. We show for the first time that the direct detection experiments' sensitivities have already reached the exciting parameter space of these models.
We also present a parameter region where the model predictions can explain the excess in the XENON1T ionization result.

\bigskip

This work is supported by NSFC under Grant No. 11975134, the National Key Research and Development Program of China under Grant No.2017YFA0402204 and Tsinghua University Initiative Scientific Research Program.

\end{document}